\documentclass[10pt]{article}
\usepackage{graphicx}
\usepackage{amsmath}

\newtheorem{theorem}{Theorem}
\newtheorem{acknowledgement}[theorem]{Acknowledgement}

\input{tcilatex}

\begin{document}

\title{Relativistic Invariance of the Phase of a Spherical Wave, relativistic
Doppler Formula and Poincar\'{e}'s expansion of space.\\
}
\author{Dr. Yves Pierseaux (ypiersea@ulb.ac.be)}
\maketitle

\begin{abstract}
Recently Einstein's invariance of the phase of a plane wave (1905) has been
described as ''questionable'' \cite{1}. Another definition of this phase,
taking into account a ''relativistically induced optical anisotropy'' for
isotropic medium in moving, has been proposed \cite{2}. We suggest
(logically) to determine this ''relativistically induced effect'' if the
isotropic medium is the \textit{vacuum}. We prove that the basic Lorentz
invariant, in vacuum, is not the phase of a \textit{plane wave (}$\omega t-%
\mathbf{\vec{k}.\vec{r})}$\textit{\ but} the phase of\textit{\ a spherical
wave (}$\omega t-kr\mathbf{)}$. According to Poincar\'{e}, an isotropic
spherical wave is not LTed (Lorentz transformed) into an isotropic spherical
wave (Einstein 1905) but LTed into \textit{an anisotropic ellipso\"{i}dal
wave} ($c=1$, relativity of simultaneity). Poincar\'{e}'s ellipsoidal
wavefront (1906) is an \textit{equiphase} surface. Our approach consists in
deleting ($\mathbf{\vec{k}.\vec{r}\rightarrow }$ $kr)$ and not adding ($%
\omega \rightarrow \mathbf{\vec{k}.\vec{u}}$) a scalar product \cite{2}. The
Lorenz gauge is connected with the invariance of the phase of a spherical
wave and the transverse gauge with Einstein's invariant. From Poincar\'{e}'s
invariant we deduce a Doppler formula that is not the same as Einstein's
one. We call this formula ''Poincar\'{e}'s formula'' because it is
unseparable of Poincar\'{e}'s theory of expansion of space and therefore the
measurements of Hubble.
\end{abstract}

\section{Invariance of the phase of a plane wave, transverse gauge and
Einstein's spherical wavefronts}

\bigskip Einstein admits that the Galilean invariant, the phase of a
monochromatic plane wave, $\phi =\omega t-\mathbf{\vec{k}.\vec{r}}$, is 
\textit{by definition} a Lorentz invariant, $\omega ^{\prime }t^{\prime }-%
\mathbf{\vec{k}}^{\prime }.\mathbf{\vec{r}}^{\prime }$, in frames in uniform
translation K and K' \cite[paragraphe 7]{3} ($k=\frac{2\pi }{\lambda }=2\pi
\nu =\omega $, $k^{\prime }=\frac{2\pi }{\lambda ^{\prime }}=2\pi \nu
^{\prime }=\omega ^{\prime }$): 
\begin{equation}
\phi =\omega t-\mathbf{\vec{k}{\Huge .}\vec{r}=}2\pi \nu (t-x\cos \theta
-y\sin \theta )\qquad =\qquad \phi ^{\prime }=\omega ^{\prime }t^{\prime }-%
\mathbf{\vec{k}}^{\prime }{\Huge .}\mathbf{\vec{r}}^{\prime }=2\pi \nu
^{\prime }(t^{\prime }-x^{\prime }\cos \theta ^{\prime }-y^{\prime }\sin
\theta ^{\prime })
\end{equation}
The invariance (1) implies that the unprimed plane wavefront, defined by the 
\textit{scalar product} $\mathbf{\vec{k}.\vec{r}}=c\mathbf{,}$ is\textit{\
not only} orthogonal ($\varphi =90%
{{}^\circ}%
)$ to the direction of propagation $\mathbf{\vec{k}}$ in K\textit{\ but also}
that the primed plane wavefront, defined by the \textit{scalar product} $%
\mathbf{\vec{k}}^{\prime }\mathbf{.\vec{r}}^{\prime }=c^{\prime },$ is
orthogonal ($\varphi ^{\prime }=90%
{{}^\circ}%
)$ to the direction of propagation $\mathbf{\vec{k}}^{\prime }$ in K'. We
have therefore $\mathbf{\vec{A}}=\mathbf{\vec{A}}_{0}\sin \phi $ and $%
\mathbf{\vec{A}}^{\prime }=\mathbf{\vec{A}}_{0}^{\prime }\sin \phi ^{\prime
} $ where $\mathbf{\vec{A}}$ and $\mathbf{\vec{A}}^{\prime }$ is
respectively the amplitude in K and K'(at two space dimensions, like
Einstein):

\begin{equation}
\mathbf{\vec{A}}\perp \mathbf{\vec{k}}\ \rightarrow \mathbf{\vec{A}}^{\prime
}\perp \mathbf{\vec{k}}^{\prime }\text{\ \ \ \ or\ \ }\mathbf{\vec{A}}.%
\mathbf{\vec{k}=\vec{A}}^{\prime }.\mathbf{\vec{k}}^{\prime }=0=Ak\cos
\varphi =A^{\prime }k^{\prime }\cos \varphi ^{\prime }\text{ \ \ \ \ }%
\Leftrightarrow \text{ \ \ \ }\varphi =\varphi ^{\prime }=90%
{{}^\circ}%
\end{equation}
That is true if we have $``\mathbf{\vec{A}=\vec{E}"}$ (Electric field), $``%
\mathbf{\vec{A}=\vec{H}"}$(Magnetic field) \textit{but also} if \ $``\mathbf{%
\vec{A}=\vec{A}"}$ where $\mathbf{\vec{A}}$ is the \textit{potential vector}%
. With the invariance of (1) Einstein therefore \textit{implicitly} adopts%
\footnote{%
Einstein defines explicitely the normality in K \textit{and in K'} : {\small %
''If we call the angle }$\theta ^{\prime }$\textit{\textit{{\small \ in K'
the angle between the wave-normal (the direction of the ray) and the
direction of moving'' } }}\cite[paragraphe 7]{3}.} the \textit{transverse
electromagnetic gauge for the potential vector}: 
\begin{equation}
div\mathbf{\vec{A}}=div\mathbf{\vec{A}}^{\prime }=0
\end{equation}

We showed that the Lorentz transformation (LT with $\gamma =(1-\beta ^{2})^{-%
\frac{1}{2}}$ and Poincar\'{e}'s units of \ space-time $"c=1"$)

\begin{equation}
x^{\prime }=\gamma (x-\beta t)\qquad \qquad \qquad y^{\prime }=y\qquad
\qquad (z^{\prime }=z)\qquad \qquad t^{\prime }=\gamma (t-\beta x)
\end{equation}
modifies not only the direction of propagation ($\mathbf{\vec{k}}\rightarrow 
\mathbf{\vec{k}}^{\prime })$ \textit{but also} the right angle\textit{\ }$%
\varphi $\textit{\ }$\rightarrow $ $\varphi ^{\prime }\neq 90%
{{}^\circ}%
$ (see 8b) \cite{4}. Einstein's invariant (1) is a Galilean invariant
because the time $t$ is constant on the wavefront in K and the Lorentz
transformed (LTed) time $t^{\prime }$ is \textit{also constant }on the
wavefront in K'. Einstein's theory of wavefront (set of simultaneous of
events) is entirely consistent with Einstein's spherical waves in the
fundamental kinematics part of his work \cite[paragraph 3]{3}:{\small \ ''At
the time }$t=\tau =0${\small , when the origin of the two coordinates (K and
k) is common to the two systems, let a \textbf{spherical wave} be emitted
therefrom, and be propagated -with the velocity c in system K. If x, y, z be
a point just attained by this wave, then }$x^{2}+y^{2}+z^{2}=c^{2}t.${\small %
\ Transforming this equation with our equations of transformation, we obtain
after a simple calculation }$\xi ^{2}+\eta ^{2}+\zeta ^{2}=c^{2}\tau ^{2}.$%
{\small \ The wave under consideration is therefore no less a \textbf{%
spherical wave}\ with velocity of propagation c when viewed in the moving
system k.''.} According to Einstein, the spherical (isotropic) \textit{shape 
}of the wavefront in K is LTed into a spherical (isotropic) shape of the
wavefront in K'. Given that Einstein's plane wavefront (1) is tangent to
Einstein's spherical wavefront within each system K and K', Einstein's
theory of wavefronts (plane or spherical) is therefore inseparable, in
optics and in kinematics as well, from the transverse gauge (3).

If the orthogonal plane $\mathbf{\vec{k}.\vec{r}}$ is transformed into an
orthogonal plane $\mathbf{\vec{k}}^{\prime }\mathbf{.\vec{r}}^{\prime }$
tangent to Einstein's primed sphere, Huang's negative frequencies are
impossible because $\mathbf{\vec{k}}^{\prime }\mathbf{.\vec{r}}^{\prime }>0$%
. But if $\ \mathbf{\vec{k}.\vec{r}}$\ \ is \textit{LTed} into a
non-orthogonal plane (\S 2), negative $\omega ^{\prime }<0$ Huang's
frequencies $\mathbf{\vec{k}}^{\prime }\mathbf{.\vec{r}}^{\prime }=\omega
^{\prime }t^{\prime }<0$ ($t^{\prime }$ is constant) become possible \cite{1}%
. This is the reason why Gjurchinovski have to introduce a
''relativistically-induced optical anisotropy''\cite{2}. The problem is
that, in Einstein's kinematics, induced by Einstein's interpretation of
invariance of light velocity with spherical wavefronts, the\textit{\
anisotropy} does not appear. In the second part (electrodynamic and optic
applications), Einstein deduces a formula for Doppler effect\footnote{%
Given that we have $(1)\Longrightarrow (5),$ if we modify (1) ($\omega
\rightarrow $ $\mathbf{\vec{k}.\vec{u}}$) for isotropic medium including
vacuum \cite{2}, how can we modify Einstein's relativistic Doppler formula
by taking into account anisotropy? (14).} (\textbf{5c}), from the invariance
(1) with LT (4),

\begin{equation}
\qquad \phi =\phi ^{\prime }\qquad \qquad \Longrightarrow \qquad \frac{%
\omega ^{\prime }}{\omega }=(\frac{\nu ^{\prime }}{\nu })_{EINSTEIN}=\gamma
(1-\beta \cos \theta )\text{ \ \ }
\end{equation}
in coupling with formulas for stellar aberration effect, $\cos \theta
^{\prime }=\frac{\cos \theta -\beta }{1-\beta \cos \theta }$ and\ $\sin
\theta ^{\prime }=\frac{\sin \theta \text{ }}{\gamma (1-\beta \cos \theta )}$%
,\ with the source $(\theta )$ at the infinity in K and the moving observer $%
(\theta ^{\prime })$ in O' in K'\cite[paragraph 7]{3}.

\section{Poincar\'{e}'s ellipsoidal wavefront, invariance of the phase of a
spherical wave and Lorenz gauge}

\bigskip Poincar\'{e} introduces the elongated ellipsoid in 1906: {\small %
''Imagine an observer and a source involved together in the transposition.
The wave surfaces emanating for the source will be \textbf{spheres}, having
as centre the successive positions of the source. The distance of this
centre from the present position of the source will be proportional to the
time elapsed since the emission - that is to say, to the radius of the
sphere. But for our observer, on account of the contraction, all these
spheres will appear as \textbf{elongated ellipsoids}. The compensation is
now exact, and this is explained by Michelson's experiments.''} \cite{5}. So
we don't need here Poincar\'{e}'s interpretation,\ very special and
unexpected, of Lorentz contraction (conclusion) but only the ''LT of a
spherical wave''. Suppose that a source S, at rest O in K, emits a circular
wavefront in $t^{\prime }=t=0$ when O and O' coincide (we\ first work in two
dimensions). \ Let us consider the relativistic invariant (light velocity $%
"c=1"$, (4), is obviously an invariant): 
\begin{equation}
x^{2}+y^{2}=r_{1}^{2}=t_{1}^{2}\qquad (a)\qquad \qquad x^{\prime
2}+y^{^{\prime }2}=r^{\prime 2}=t^{\prime 2}\qquad (b)
\end{equation}
The shape of (6a) is by definition a circle in K ($r=r_{1}=t=t_{1})$. What
is the LTed shape in K' of (6b)? If the time t is fixed in K the time 
\textit{t'} is not fixed (by LT) \textit{in} K'. The LTed shape of 6(b) with
(4) $t^{\prime }=\gamma ^{-1}t_{0}-\beta x^{\prime }$ is: 
\begin{equation}
x^{\prime 2}+y^{^{\prime }2}=(\gamma ^{-1}t_{1}-\beta x^{\prime })^{2}\qquad
or\qquad (\gamma ^{-1}x^{\prime }+\beta t_{1})^{2}+y^{\prime 2}=t_{1}^{2}
\end{equation}
It is the Cartesian equation of an \textit{elongated ellipse}\ in K' with
the observer O' at the focus F. The physical meaning of Poincar\'{e}'s
ellipse $"c=1"$ is the relativity of simultaneity: two simultaneous events
in K are not simultaneous in K'. In polar coordinates the equation of the
ellipse ($\theta ^{\prime }$ is the polar angle with F as pole, $\beta $ is
eccentricity and $\gamma t_{1}$ the large axis) is $(r,\theta
)_{K-cercle(c=1)}\rightarrow (r^{\prime },\theta ^{\prime })_{K^{\prime
}-ellipse(c=1)}$ :

\begin{equation}
r^{\prime }=\frac{r_{0}}{\gamma (1+\beta \cos \theta ^{\prime })}=t^{\prime }%
\text{ \ \ }(a)\qquad \qquad \qquad \tan \varphi ^{\prime }=\frac{\beta \cos
\theta ^{\prime }+1}{\beta \sin \theta ^{\prime }}\text{ \ \ \ }(b)
\end{equation}
We showed that a tangent to the circle is LTed into a (non-orthogonal)
tangent to the ellipse (8a) that makes the angle $\varphi ^{\prime }$ with
the direction of propagation $\mathbf{\vec{k}}^{\prime }$ (8b \cite{4}, only
for longitudinal $\theta ^{\prime }=0$ we have $\varphi ^{\prime }=90%
{{}^\circ}%
$): the \textit{anisotropy }in LTed frame K' is determined not only by the
angle $\theta ^{\prime }$ \textit{but also by the angle }$\varphi ^{\prime }$
(2) for the tangent plane wave$.$ At three dimensions, according to
Poincar\'{e}, the spherical shape is LTed into an elongated ellipsoidal
shape. Poincar\'{e}'s ellipsoid of revolution is the direct kinematical
explanation of the very physical \textit{''headlight effect'' } (in
synchrotron radiation, bremsstrahlung...): the isotropic (spherical)
emission of a moving source is anisotropic (ellipsoidal) observed from a
system at rest with relativistic transformation of the solid angles $\Omega
=2\pi (1-\cos \theta )$ and $\Omega ^{\prime }=2\pi (1-\cos \theta ^{\prime
})$. The reduction of the angle of aperture of the cone of emission of a
moving source ($d\Omega ^{\prime }=d\Omega \frac{1}{\gamma ^{2}(1+\beta \cos
\theta )^{2}}$, the azimuth angle $\psi =\psi ^{\prime }$ being invariant)
is engraved in the ellipsoidal shape of the wavefront. So the question is:
if the invariant is not the \textit{shape}\ of a spherical wave, what is the
true Lorentz invariant? The answer is:\ it is the \textit{phase}\textbf{\ }$%
\Psi $ of the spherical wave:\ 

\begin{equation}
\Psi =\omega t-kr=2\pi \nu (t-r)\qquad \qquad =\qquad \Psi ^{\prime }=\omega
^{\prime }t^{\prime }-k^{\prime }r^{\prime }=2\pi \nu ^{\prime }(t^{\prime
}-r^{\prime })\ 
\end{equation}

The spherical (monochromatic) wave, $A(r,t)=A_{0}\frac{e^{i\Psi }}{r}$, is
LTed into an ellipsoidal (monochromatic) anisotropic wave, $A^{\prime
}(r,t)=A_{0}^{\prime }\frac{e^{i\Psi \prime }}{r^{\prime }}$ ($r^{\prime
}=t^{\prime }$ variable).\textit{\ }The LTed\textit{\ ellipsoidal wavefront
is an equiphase surface} (a non-transverse section in Minkowski's cone).
First place is given by Poincar\'{e} to spherical waves and by Einstein to
plane waves. Fortunately we have in optics the principle of Huygens that
enable to determine any front, including\ the plane front, on the basis of
spherical waves \cite{7} Poincar\'{e}'s introduction of the potential
four-vector ($\mathbf{A},$ $V)$ is fundamental because if the scalar
potential is zero in K ($V=0)$, it cannot be zero in K' ($V^{\prime }\neq 0)$
\cite{6}. So by the integration of potential vector $\mathbf{A}$ on the
spherical surface ($dS=r^{2}d\Omega $ $=$ $dS^{\prime }=r^{\prime 2}d\Omega
^{\prime }),$ and by application of the Gauss-Ostrogradski's theorem ($%
dV=r^{2}\sin \theta d\theta d\psi $ $=dV^{\prime }=r^{\prime 2}\sin \theta
^{\prime }d\theta ^{\prime }d\psi )$,\ the flux of $\mathbf{A}$ is zero in
K, $div\mathbf{\vec{A}}=0$ ($\mathbf{\vec{A}}$ is in a tangent plane to the
sphere $\mathbf{\vec{A}.\vec{k}}=0$), but it is\textit{\ not zero} in K': $%
div\mathbf{\vec{A}}^{\prime }=\frac{-\partial V^{\prime }}{\partial
t^{\prime }}$ ($\mathbf{\vec{A}}^{\prime }$ is in a tangent plane to the
ellipsoid K', $\mathbf{\vec{A}}^{\prime }\mathbf{.\vec{k}}^{\prime }\mathbf{%
\neq 0}$). Poincar\'{e}'s ellipsoid is therefore directly connected with
Lorenz gauge \cite{7}: 
\begin{equation}
div\mathbf{\vec{A}}+\frac{\partial V}{\partial t}=div\mathbf{\vec{A}}%
^{\prime }+\frac{\partial V^{\prime }}{\partial t^{\prime }}=0
\end{equation}
In order to cure Huang's negative frequencies, Gjurchinovski adds a dot in $%
\mathbf{\vec{k}}^{\prime }\mathbf{.\vec{u}}^{\prime }\mathbf{=\omega }%
^{\prime }$ whist we delete a dot in $\mathbf{\vec{k}}^{\prime }\mathbf{.%
\vec{r}}^{\prime }$. According to Gjurchinovski, in his Figure 3, if the
medium is isotropic, the LTed vector $\mathbf{\vec{u}}$ is \textit{%
non-orthogonal} to the LTed front but, \textit{if the isotropic medium is
the isotropic vacuum,} the LTed vector $\mathbf{\vec{u}}$ , that is aligned
onto the LTed vector wave $\mathbf{\vec{k}}$, is \textit{orthogonal} to the
LTed front. Why the electromagnetic vacuum would be the only isotropic
medium that would escape to ''relativistically induced optical anisotropy''?
The electromagnetic vacuum is a medium exactly as the others (with a
permittivity, a permeability, an impedance, 377 $\Omega $...) and therefore
we have in Lorenz gauge a non-transversal LTed wavefront. Let us note that
nothing is changed at the level of electromagnetic \textit{fields} because
Poincar\'{e}'s longitudinal component $\mathbf{\vec{A}}_{\parallel }^{\prime
}$ is compensated\footnote{%
We showed that Einstein's cancellation of the potential V' is exactly
''Einstein's deletion of aether''\cite{4}. In Poincar\'{e}'s Lorenz gauge, $%
V^{\prime }=A_{\parallel }^{\prime }\neq 0,$ we have a relativistic aether
like, for example, the thermodynamical cosmic background radiation (CBR),
with respect to which one can only measures relative velocities.} \cite{8}:

\begin{equation*}
\mathbf{\vec{E}}_{\parallel }^{\prime }=-\partial _{t}\mathbf{\vec{A}}%
_{\parallel }^{\prime }-\nabla V^{\prime }=0
\end{equation*}
We underline that, according to Poincar\'{e}, the wave vector $\mathbf{\vec{k%
}}^{\prime }$, the transversal electric field $\mathbf{\vec{E}}^{\prime }$
and the transversal magnetic field $\mathbf{\vec{H}}^{\prime }$ define a
trirectangle trihedron: the non-transversality is only at the level of the
potential vector $\mathbf{\vec{A}}^{\prime },$ that, according to ''the main
stream'' is a \textit{philosophical} entity: a gauge effect would be
absolutely impossible! Maybe... but the use a relativistic gauge (10) in a
relativistic theory is a good idea and we are moreover able to deduce from
Poincar\'{e}'s invariant (9) a Doppler formula, taking into account a
fundamental anisotropy (14). In order to dot that, we have to find the LT of 
$r$ into $r^{\prime }$ in (9) and to introduce the angles $\theta $ and $%
\theta ^{\prime }$. From (4) we have the defined positives norms $r=\sqrt{%
x^{2}+y^{2}}$ and $r^{\prime }=\sqrt{x^{\prime 2}+y^{\prime 2}}$ and in
polar coordinates, the relativistic law of composition of velocities is
(with $\frac{r}{t}=v,$ $\frac{r^{\prime }}{t^{\prime }}=v^{\prime }$, $%
v_{x}=v\cos \theta ,$ $v_{y}=v\cos \theta $, $v_{x^{\prime }}^{\prime
}=v^{\prime }\cos \theta ^{\prime },$ $v_{y^{\prime }}^{\prime }=v^{\prime
}\cos \theta ^{\prime }$)$:$

\begin{equation}
v^{\prime }\cos \theta ^{\prime }=\frac{v\cos \theta -\beta }{1-\beta v\cos
\theta }\qquad (a)\qquad v^{\prime }\sin \theta ^{\prime }=\frac{v\sin
\theta }{\gamma (1-\beta v\cos \theta )}\qquad (b)\qquad \Leftarrow \ \ \ \
\ \ \frac{t^{\prime }}{t}=\gamma (1-\beta v\cos \theta )\qquad (c)
\end{equation}
For a light point in uniform moving (displacement under constant angle $%
\theta $) of a spherical front , with $"\frac{r}{t}=v=c=1$ and $\frac{%
r^{\prime }}{t^{\prime }}=v^{\prime }=c^{\prime }=1"$ we have:

\begin{equation}
\cos \theta ^{\prime }=\frac{\cos \theta -\beta }{1-\beta \cos \theta }\text{
\ \ }(a)\text{\ \ \ \ }\sin \theta ^{\prime }=\frac{\sin \theta }{\gamma
(1-\beta \cos \theta )}\text{ \ \ }(b)\text{\ \ }\Leftarrow \text{\qquad }%
\frac{t^{\prime }}{t}=\gamma (1-\beta \cos \theta )=\frac{r^{\prime }}{r}%
\qquad (c)
\end{equation}
So with 12(c) and from Poincar\'{e}'s invariant (9) we deduce, exactly like
Einstein, two coupled formulas for the relativistic transformation of angle
and the relativistic transformation of frequency:

\begin{equation}
\frac{\omega ^{\prime }}{\omega }=\frac{\nu ^{\prime }}{\nu }=\frac{1}{%
\gamma (1-\beta \cos \theta )}=\gamma (1+\beta \cos \theta ^{\prime })\text{
\ \ }(a)\text{\qquad \qquad }\frac{r^{\prime }}{r}=\frac{1}{\gamma (1+\beta
\cos \theta ^{\prime })}\text{ \ }(b)
\end{equation}
With (13b) we find again the ellipse (8a). In order to have a true Doppler
formula, the observer in K' have to receive the signals under constant angle 
$\theta ^{\prime }$ (stellar aberration) and with constant frequency $\nu
^{\prime }$: \textit{at the infinity} \textit{Poincar\'{e}'s spherical waves
become Poincar\'{e}'s plane wave}, with constant angle $\varphi ^{\prime }$
(8b). In the same way as Einstein's plane wave, we have to \textit{define a
infinite distance }between source (in O) and the observer (in K'). So in
order to compare (13a) with Einstein's model (5) we have to change the sense
(the sign) of light velocity (also in 9, see \cite{7}, principle of inverse
return of the light): 
\begin{equation}
\Psi ^{+}=\Psi ^{\prime +}\qquad \Longrightarrow \qquad \qquad (\frac{\omega
^{\prime }}{\omega })_{\infty }=(\frac{\nu ^{\prime }}{\nu }%
)_{POINCARE}=\gamma (1-\beta \cos \theta ^{\prime })
\end{equation}
The observer ($\theta ^{\prime })$, Einstein or Poincar\'{e}, and the source
($\theta )$ are now in identical configuration for (5) and (14). We suggest
to call (14) ''Poincar\'{e}'s'' formula for the same reason for which
Poincar\'{e} had called (4) the ''Lorentz''' transformation. We underline
the following contrast: Poincar\'{e}'s anisotropy (transformation of angles,
headlight effect) has a physical meaning \textit{before} the limit at the
infinity (stellar aberration) but (13) only becomes a Doppler formula 
\textit{after} the limit at the infinity. In (13), we have one only source
in O and an infinity of observers, in moving, onto a finite ellipsoid. By
inverting (13) into (14)$_{\infty }$ we have an infinity of sources, in
moving, onto an \textit{infinite ellipsoid} and one only (terrestrial)
observer at the focus of this ellipsoid \cite[fig 2]{7}. Moreover in
non-relativistic optics we can define a Doppler formula from spherical
waves, without a limit at the infinity (one considers that the source,
velocity v, is ''almost'' at rest during two successive fronts, velocity c).
Non-relativistic hypothesis v\TEXTsymbol{<}\TEXTsymbol{<}c being
unacceptable for a relativistic theory, everything happens therefore as if
Poincar\'{e}'s formula should have to be applied to (very) remote objects.

\section{Poincar\'{e}-Hubble's expansion of space and the relativistic
Doppler formula}

Poincar\'{e}'s formula and Einstein's formula cannot be the same because, in
the former (5), the frequency is LTed as the time whilst, in the latter
(14), the frequency is LTed as the \textit{inverse }of the time (i.e., the
lengthwave is LTed as Poincar\'{e}'s distance \cite{4}). Einstein's
four-vector of (plane) wave ($\mathbf{\vec{k}},$ $\omega )$ and
Poincar\'{e}'s potential four-vector ($\mathbf{\vec{A}},$ $V)$ of spherical
electromagnetic wave are antinomic. The following contrast is therefore
irreducible:

\begin{equation}
(\frac{\nu ^{\prime }}{\nu })_{EINSTEIN}=\gamma (1-\beta \cos \theta )\qquad
\qquad \qquad (\frac{\nu ^{\prime }}{\nu })_{POINCARE}=\gamma (1-\beta \cos
\theta ^{\prime })
\end{equation}
The polemical question of historical priorities, between Poincar\'{e}'s
supporters and Einstein's supporters, for the theory of relativity is 
\textit{irrelevant} because it is \textit{not} the same theory. The dilation
factor is in both formulas but the respective role of the angle of the source%
\textit{\ }$\theta $ et $\theta ^{\prime }$\textit{\ are reverted}. Most of
the experimental tests of relativistic Doppler formulas have been realized
in the longitudinal configuration (\cite{9} \& \cite{10}) or for very high
velocities, where both formulas are identical: 
\begin{equation}
(\frac{\nu ^{\prime }}{\nu })_{EINSTEIN-POINCARE}=\sqrt{\frac{1-\beta }{%
1+\beta }=}\gamma (1-\beta )
\end{equation}
. Hasselkamp's experimentation, in (almost) transversal configuration \cite
{11} \textit{seems to plead} for Einstein's formula, but the situation is
not the same if we define the transversality with the angle of observer or
with the angle of the source. For $\theta =90%
{{}^\circ}%
$ \ the transversal effect predicted by Einstein is a contraction of
duration while this one predicted by Poincar\'{e}'s formula is a dilation of
duration. It is the inverse if we take $\theta ^{\prime }=90%
{{}^\circ}%
.$ There is no purely terrestrial experiment already realized in order to
decide \textit{clearly }between the two formulas (15). We underline that
Einstein deduced his Doppler formula in the second part of his famous work
(electrodynamics application of his kinematics) on the basis of an
additional hypothesis (1). On the other hand, our deduction of Poincar\'{e}%
's formula (14) is intrinsically engraved in LT (4) and therefore it becomes
impossible to distinguish optics and kinematics (structure of space-time). 

So only one cosmic experiment \textit{seems to plead} for Poincar\'{e}'s
Doppler formula, which is inseparable from his theory of the \textit{%
expansion of space }(according to Poincar\'{e} the Lorentz contraction of 
\textit{units of measure} induces a \textbf{dilated} \textbf{distance},
measured in ''light-year'', see the quotation! and \cite{12}): this is the
experiment of Hubble (1929). 

On the other hand, Einstein-de Broglie's plane wave (1) is a solution of Schr%
\"{o}dinger equation, \textit{that is not a relativistic equation}. This is
the reason why quantum atomic structure should be not concerned by Poincar%
\'{e}'s relativistic expansion: It is impossible to make physics without
''Einstein's quantum principle of identical units of measure'' \cite{13}. A
synthesis between Einstein's and Poincar\'{e}'s kinematics that should go
into action \textit{not} at the same scale, is therefore not only possible
and but even necessary (i.e a synthesis between Einstein's photons ($\mathbf{%
\vec{k}},$ $\omega )$ and Poincar\'{e}'s electromagnetic waves ($\mathbf{%
\vec{A}},$ $V)$ , see note 3 and \cite{14}),.

\begin{acknowledgement}
Je d\'{e}die cette lettre \`{a} Jean Reignier gr\^{a}ce \`{a} qui j'ai pu
commencer il y a vingt ans des recherches sur la ''structure fine'' de la
relativit\'{e} restreinte (Einstein et Poincar\'{e}): il y a deux th\'{e}%
ories tr\`{e}s proches et n\'{e}anmoins distinctes (15) \cite{15}. Je
remercie \'{e}galement Pierre Marage, Nicolas van steenkiste et Germain
Rousseaux. Ce dernier attir\'{e} mon attention sur l'existence d'une
''structure fine'' de l'\'{e}lectromagn\'{e}tisme classique et donc sur la th%
\'{e}orie de Riemann-Lorenz\cite{16}.
\end{acknowledgement}

\end{document}